\documentclass[conference]{IEEEtran}
\IEEEoverridecommandlockouts
\newif\ifdoubleblind
\doubleblindfalse  % Camera-ready version

\usepackage{balance}
\usepackage{multirow}
\usepackage{tikz}
\usepackage{adjustbox}
\usepackage{amsmath,amssymb,amsfonts}
\usepackage{algorithmic}
\usepackage{graphicx}
\usepackage{numprint}
\usepackage{siunitx}  % For decimal alignment in percentages
\usepackage[inline]{enumitem}
\usepackage{booktabs}
\npdecimalsign{.}
\usepackage{textcomp}
\usepackage[compatibility=false]{caption}  % Provides better caption handling
\usepackage{subcaption}
\usepackage{listings}
\lstdefinestyle{mystyle}{
    language=C++,                   % Set the programming language
    basicstyle=\small\ttfamily,  % Match ACM body text size with typewriter font
    keywordstyle=\color{blue},      % Set keyword color
    commentstyle=\color{gray},      % Set comment color
    numbers=left,                   % Display line numbers
    numberstyle=\footnotesize,      % Match line number size to code font
    numbersep=5pt,                  % Separation between line numbers and code
    breaklines=true,                % Enable line wrapping
    frame=single,                   % Add a frame around the code
    tabsize=4                       % Set the tab size
}

\lstdefinestyle{prompt}{
    language=C++,              % Set the programming language
    basicstyle=\ttfamily,      % Set the font style
    keywordstyle=\color{blue}, % Set keyword color
    commentstyle=\color{green},% Set comment color
    numbers=none,              % Display line numbers
    numberstyle=\tiny,         % Set the style for line numbers
    numbersep=5pt,             % Set the separation between line numbers and code
    breaklines=true,           % Enable line wrapping
    frame=single,              % Add a frame around the code
    tabsize=4                  % Set the tab size
}

\lstdefinestyle{CStyle}{
    backgroundcolor=\color{backgroundColour},   
    commentstyle=\color{mGreen},
    keywordstyle=\color{magenta},
    numberstyle=\tiny\color{mGray},
    stringstyle=\color{mPurple},
    basicstyle=\footnotesize,
    breakatwhitespace=false,         
    breaklines=true,                 
    captionpos=b,                    
    keepspaces=true,                 
    numbers=left,                    
    numbersep=5pt,                  
    showspaces=false,                
    showstringspaces=false,
    showtabs=false,                  
    tabsize=2,
    language=C
}
\usepackage[framemethod=tikz]{mdframed}
\mdfdefinestyle{mpdframe}{
    frametitlebackgroundcolor = black!15,
    frametitlerule            = true,
    roundcorner               = 8pt,
    middlelinewidth           = 1pt,
    innermargin               = 0.25cm,
    outermargin               = 0.25cm,
    innerleftmargin           = 0.25cm,
    innerrightmargin          = 0.25cm,
    innertopmargin            = 2mm,
    innerbottommargin         = 2mm,
    skipabove                 = 6pt,
    skipbelow                 = 12pt,
}

\mdfdefinestyle{insight}{%
    style=mpdframe,
    frametitle={Insight},
}
    
\usepackage{hyperref}
\usepackage{comment}
\definecolor{dkgreen}{rgb}{0,0.6,0}
\definecolor{gray}{rgb}{0.5,0.5,0.5}
\definecolor{mauve}{rgb}{0.58,0,0.82}
\newcommand{\ericsson}{our-org-(redacted) }

\title{Auto-repair without test cases: How LLMs fix compilation errors in large industrial embedded code\\
\footnotesize \textsuperscript{}
\ifdoubleblind
\else
\thanks{This work was partially supported by the Wallenberg Artificial Intelligence, Autonomous Systems and Software Program (WASP) funded by the Knut and Alice Wallenberg Foundation.}
\fi
}
\makeatletter

\newcommand{\linebreakand}{%
  \end{@IEEEauthorhalign}
  \hfill\mbox{}\par
  \mbox{}\hfill\begin{@IEEEauthorhalign}
}
\makeatother

\ifdoubleblind
  \author{\IEEEauthorblockN{Anonymous Authors}}
\else
\author
{\IEEEauthorblockN{Han Fu\IEEEauthorrefmark{1}\IEEEauthorrefmark{2},
Sigrid Eldh\IEEEauthorrefmark{1}\IEEEauthorrefmark{3},
Kristian Wiklund\IEEEauthorrefmark{1},
Andreas Ermedahl\IEEEauthorrefmark{1}\IEEEauthorrefmark{2},
Philipp Haller\IEEEauthorrefmark{2} and
Cyrille Artho\IEEEauthorrefmark{2}}
\IEEEauthorblockA{\IEEEauthorrefmark{1}\textit{Ericsson AB,}
Stockholm, Sweden \\
Email: \{han.fu,\,sigrid.eldh,\,kristian.wiklund,\,andreas.ermedahl\}@ericsson.com
\IEEEauthorblockA{\IEEEauthorrefmark{2}
\textit{KTH Royal Institute of Technology,}
Stockholm, Sweden \\
Email: \{phaller,\,artho\}@kth.se
}
\IEEEauthorblockA{\IEEEauthorrefmark{3}
\textit{Mälardalen University,}
Västerås, Sweden\\
}}}
\fi

\begin{document}
\maketitle

\begin{abstract}
The co-development of hardware and software in industrial embedded systems frequently leads to compilation errors during continuous integration (CI). Automated repair of such failures is promising, but existing techniques rely on test cases, which are not available for non-compilable code.

We employ an automated repair approach for compilation errors driven by large language models (LLMs). Our study encompasses the collection of more than~\numprint{40000} commits from the product's source code.
We assess the performance of an industrial CI system enhanced by four state-of-the-art LLMs, comparing their outcomes with manual corrections provided by human programmers. LLM-equipped CI systems can resolve up to 63\,\% of the compilation errors in our baseline dataset. Among the fixes associated with successful CI builds, 83\,\% are deemed reasonable. Moreover, LLMs significantly reduce debugging time, with the majority of successful cases completed within 8 minutes, compared to hours typically required for manual debugging. 
\end{abstract}


\section{Introduction}\label{introduction}
Hardware/software co-development in the context of industrial embedded systems frequently results in build failures during the continuous integration (CI) phase, notably within the compilation step. Build failures are common in large organizations and are often caused by dependency issues~\cite{fu2022prevalence}.

These dependency issues cannot be resolved easily, as the incompatibilities stem from developers' slightly different setups, caused by the misalignment in parallel development between hardware and software~\cite{fu2022prevalence,FuEWEHA24}. A previous paper~\cite{FuEWEHA24} reports that dependency issues, including static check failures, account for 76\,\% of the 14 recognized types of compilation errors, with the majority of solutions involving modifications spanning 1--4 lines. Motivated by these findings, our approach focuses on dependency-related compilation errors in an industrial CI system, as resolving build failures is a critical first step before testing.

Integrating automatic fixing methods into CI workflows is essential to effectively address this challenge. However, traditional automatic program repair (APR) techniques often encounter challenges with compilation errors and producing high-quality patches necessitates a robust test suite~\cite{DBLP:conf/icse/Motwani21}. 

While automatic repair methods aim to reduce human effort, generating complex fixes remains difficult~\cite{dikici2025advancements}. Large language models (LLMs) excel in generating human-like output, supporting various applications like question-answering and code generation systems~\cite{DBLP:journals/corr/abs-2307-06435}. Likewise, many bug prediction and repair methods rely heavily on machine learning algorithms~\cite{jin2023inferfix}. 

We investigate the following research questions:

\textbf{RQ1}~\textit{Can an LLM generate useful fixes for compilation errors?}\label{RQ1}

\textbf{RQ2}~\textit{What type of LLMs and prompts setup are effective for addressing compilation errors?}\label{RQ2}

\textbf{RQ3}~\textit{What efficiency can be expected of an LLM-equipped CI build pipeline for accurately and automatically resolving compilation errors?}\label{RQ3}

To answer these questions, we design and evaluate an LLM-based approach for automatic compilation error repair in an industrial CI environment. Through our experiments, we demonstrate that LLMs provide substantial support in this setting: our system achieves a 63\,\% CI pass rate on previously failing builds, with 83\,\% of successful fixes confirmed as accurate by developers. This indicates that LLM-assisted repair can significantly improve debugging efficiency in practice.

LLMs need the right amount of context for their fixes, but such a context is scarce for compilation errors. Our automated repair solution for CI systems addresses this limitation by enhancing the compilation error information with fix templates that guide the LLM, while tailoring prompt features to leverage the unique strengths of each LLM. The design and effectiveness of this approach are detailed in Section~\ref{sec:methodology}, where we demonstrate how these strategies help overcome the context limitations inherent in CI-based compilation errors.

Our evaluation indicates how a fine-tuned prompt significantly enhances the robustness and efficiency of automated repair CI solutions. Based on over~\numprint{40000} builds, we randomly selected around~\numprint{1000} failing builds with compilation errors, and passed them to four LLMs. Our findings are as follows:

\begin{enumerate} 
    \item We achieve a pass rate of 63\,\% in a complex industrial compilation CI environment. Our results underscore the significant potential of LLM-based automatic program repair (APR) for compilation errors in embedded systems.
    \item We conducted a prompt study based on multiple input features and strategically selected four model families—CodeT5+, CodeLlama, Falcon, and Bloom—based on four principal criteria. As a result, we were able to increase the pass rate from 8--19\,\% to up to 63\,\%.
    \item In an evaluation of randomly selected fixes, two experienced developers agree that 83\,\% of the fixes associated with successful CI builds are reasonable or even exact matches. Thus, using our method corresponds to significant time savings in debugging.
\end{enumerate}

The remainder of this paper is structured as follows: Section~\ref{sec:Background} presents the context of our research and motivates our work. Section~\ref{sec:related-work} provides an overview of related work. Section~\ref{sec:methodology} details our methodology and study design. Section~\ref{sec:studyresults} presents the results of our study, followed by a discussion in Section~\ref{sec:discussion}. Section~\ref{sec:conclusion} concludes and describes future work.

\section{Background}\label{sec:Background}
Continuous Integration (CI) is a well-established practice that automates merging, builds, and tests to support frequent, small contributions from multiple developers, providing rapid feedback and reducing error correction costs~\cite{duvall2007continuous, planning2002economic}. Despite successful local compilation and testing, developers frequently encounter dependency issues during centralized compilation~\cite{mesbah2019deepdelta}. These challenges are caused by the asynchronous development of hardware and software, driven by rapid and demanding delivery cycles. Following a previous study~\cite{FuEWEHA24}, we integrated a~\emph{Shadow Job} into the CI system, designed to autonomously analyze and fix compilation errors alongside ongoing CI processes. Its scalable design allows for easy extension to meet the evolving needs of dynamic development environments.

\subsection{Industrial continuous integration context}
\begin{figure}
    \centering  
    \includegraphics[scale=.9]{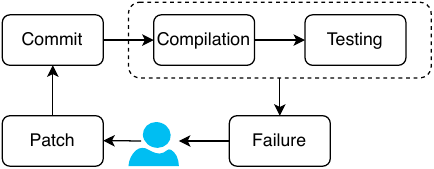}
    \caption{Industrial CI pipeline}
    \label{fig:IndustrialCIpipeline}
\end{figure}

Fig.~\ref{fig:IndustrialCIpipeline} shows an industrial CI pipeline, which consists of multiple stages executed as a cohesive atomic unit. The pipeline is divided into distinct stages: commit, compilation, and testing.
Each phase represents a set of related tools; e.\,g., compilation includes various static checks.
When a compilation or test failure occurs, the pipeline is stopped, and the error is reported to the developer for resolution. The complexity of the pipeline poses significant challenges in integrating new technologies. Any new tool must integrate smoothly with existing program analysis workflows across all CI builds, introducing potential points of failure and further increasing system complexity. The CI system examined in this study is designed to support software written in C/C++.

The CI system in our study supports a wide range of tasks that ensure product quality and stability. This results in a highly complex setup, comprising 10 to 30 stages, with each stage invoking at least 30 different software tools. The layered structure and extensive toolchain reflect the rigorous demands of modern industrial software development pipelines.

\subsection{Compilation errors in industrial CI}
The hardware team focuses on configuring and preparing hardware for deployment within the CI environment to support multiple development teams. In contrast, the software team relies on simulators or prototypes for local development, compilation, and testing in the early stages~\cite{fu2022prevalence}. This misalignment introduces significant challenges that must be addressed.

Fixing compilation errors is often more challenging than fixing test failures, as it relies solely on compiler feedback and lacks the contextual cues provided by test outcomes. While compilers effectively detect common programming errors, they often fail to accurately pinpoint error locations~\cite{DBLP:journals/ahci/Traver10}. Consequently, a novel approach is needed to address integration challenges and mitigate risks when incorporating APR tools.

In addition to the commit in Fig.~\ref{fig:IndustrialCIpipeline}, each commit or patch requires an additional code review, demanding substantial manual effort. This process is time-intensive and often delays the integration of critical fixes. To streamline this workflow, we propose an automated solution for efficiently resolving compilation errors by introducing LLMs. This approach avoids reviewing changes that do not compile and thus minimizes the manual effort needed for code review and patch generation, streamlining the integration of critical fixes.

\section{Related work}\label{sec:related-work}
Automatic Program Repair (APR) aims to automatically fix software bugs and has emerged as a significant research area in computer science. Classic APR techniques can be mainly classified as heuristic-based~\cite{le2016history,10.1145/3180155.3180233}, constraint-based~\cite{le2017s3,long2015staged,mechtaev2016angelix} and template-based~\cite{ghanbari2019practical,liu2019tbar,martinez2016astor} ones. Due to the complexity of various application scenarios, selecting the appropriate APR technique can be challenging. Additionally, existing APR techniques typically require a robust test suite to produce high-quality patches. Thus, handling compilation errors where tests cannot run is hard. We address this limitation by enabling repair before test execution.

\subsection{APR for compilation errors}
In software development, testing requires successful compilation. However, conventional repair methods struggle with compilation errors in embedded systems, and current software ecosystems lack the necessary infrastructure to support these approaches~\cite{mesbah2019deepdelta}. TRACER~\cite{ahmed2018compilation} focuses on single-line errors but struggles with multi-line and dependency issues common in real-world scenarios. Its successor, MACER~\cite{DBLP:conf/aied/ChhatbarAK20}, approaches repair as a classification task, providing detailed analyses of error types and fixes while improving training speed. DeepFix~\cite{gupta2017deepfix} addresses multi-line errors in C code but is primarily designed for student submissions, limiting its industrial relevance. We address these gaps by focusing on multi-line compilation errors in industrial contexts.

TransRepair~\cite{DBLP:conf/kbse/LiLFMXCL22} advances repair accuracy by integrating contextual features and compiler feedback. However, it emphasizes accuracy over time efficiency, a key requirement in industrial settings. Similarly, Google’s DeepDelta~\cite{mesbah2019deepdelta} achieved a 50\% patch success rate for compilation errors in industrial projects but lacks integration with complex systems and full support for automatic repair. Building on these efforts, our study combines log parsing and commit tracing techniques to tackle compilation errors in industrial CI pipelines, offering a more balanced and practical solution.

Zhang et al.~\cite{zhang2019large} conducted an extensive study of compilation errors across~\numprint{3799} open-source projects, analyzing common error types, resolution times, and reviewing 325 broken builds to identify fix patterns for the ten most frequent errors. Based on this, we analyzed over~\numprint{40000} broken builds from real production code and examined~\numprint{1000} compilation errors. Seo et al.~\cite{seo2014programmers} categorized compilation errors in centralized build systems, whereas our focus is on embedded, remote CI systems with a higher prevalence of dependency errors. Compiler error messages, often challenging to interpret~\cite{becker2019compiler}, especially for novice developers, were highlighted as a key issue by Rosen et al.~\cite{rosen1965pufft}.

\subsection{APR with Large Language Models}
LLMs have gained attention for their potential to enhance APR. Recent advancements in LLMs pre-trained on large code corpora have shown exceptional ability in understanding and generating code~\cite{DBLP:journals/corr/abs-2303-18223}, enabling accurate and context-aware fixes by capturing complex patterns and semantic structures. Despite their promise, the integration of LLMs into industrial CI systems remains limited. Transformer-based LLMs excel in various natural language processing (NLP) tasks~\cite{DBLP:journals/corr/abs-2303-18223, wei2022chain, PanLWCWW24}, but their application in industrial practices is still underexplored. This work addresses this gap by implementing LLMs into industrial CI pipelines to achieve full automation.

Xia and Zhang~\cite{xia2023keep} explore LLMs for APR, demonstrating the effectiveness in repairing multi-line errors at relatively low cost. However, the solution is conversation-based rather than offline, raising concerns about potential delays and security risks. Additionally, a conversation-based approach cannot be easily integrated into a fully automated CI workflow. Further improvements are needed to address the security and complexity challenges of industrial applications.

Moreover, LLMs developed to understand, generate, and enhance code are primarily designed to provide debugging insights without relying on compilers or interpreters~\cite{DBLP:journals/corr/abs-2307-10793}. Building on this, our work implements four state-of-the-art LLMs in an industrial CI setting to effectively address and fix compilation errors without test cases. 

\subsection{APR in CI} 
APR identifies and repairs defects with minimal human intervention~\cite{DBLP:journals/corr/abs-2307-06435}. Despite its potential, APR often faces challenges with scalability and handling diverse, complex bug types effectively~\cite{DBLP:journals/corr/abs-2303-18184}. Little research has focused on integrating APR into CI, especially in complex software systems~\cite{winter2022let} of industrial settings. Our study bridges this gap by introducing the Shadow Job approach, enabling APR integration to resolve compilation errors efficiently without disrupting workflows.

The benefits of CI include enhanced productivity and bug detection~\cite{hilton2016usage, vasilescu2015quality}, yet limited work addresses pipeline error resolution. Vassallo et al.~\cite{vassallo2017tale} identified testing, compilation, and dependency issues as common CI failures, while Beller et al.~\cite{beller2017oops} noted that most build failures stem from test errors, which are well-covered in industry~\cite{garousi2017test}. Leveraging LLMs, our method automates compilation error fixes without relying on test cases, achieving high success rates.

With proper architecture and testing, CI enables smooth integration and preserves quality. Jin et al.~\cite{jin2023inferfix} advanced LLM-based program repair by integrating it into the CI pipeline to automate software development. In contrast, our work focuses specifically on embedded systems within CI environments—an area that presents unique challenges such as higher dependency failure rates and limited observability due to remote deployment. This study highlights the need for scalable, automated solutions for compilation failures in industrial CI environments, particularly in embedded and remote systems with higher dependency error rates, a topic that remains underexplored.

\subsection{APR in industry}
Research by Naitou et al.~\cite{naitou2018toward} has highlighted the effectiveness of APR techniques in industrial software, inspiring tools like Facebook’s SapFix~\cite{marginean2019sapfix} and Getafix~\cite{bader2019getafix}. While these tools show promise across software domains, most studies focus on general-purpose development and overlook compilation errors in large-scale industrial CI systems.

\section{Methodology}\label{sec:methodology}
We conduct an empirical case study to develop and evaluate automated repair for hardware-in-the-loop CI environments. The study integrates four state-of-the-art LLMs into the repair workflow and systematically assesses their performance in resolving compilation errors in an industrial setting.

The evaluation focuses on two primary metrics: (1) the \emph{accuracy} of the generated fixes, measured by the correctness of the resulting builds and developer validation, and (2) the \emph{efficiency} of the repair process, in terms of time to resolution and reduction in manual intervention.

\subsection{Implementation}\label{sec:implementation}
We have integrated four state-of-the-art LLMs, as shown in Table~\ref{tab:LLMs}, within an \ericsson hardware-in-the-loop CI system to evaluate their efficiency and accuracy in automatically fixing compilation errors. Our architecture accounts for the challenges in integrating pre-trained LLMs for APR in a complex industrial CI process.

\subsubsection{\textbf{Shadow Job}}\label{subsubsec:ShadowJob}
\begin{figure}
    \centering  
    \includegraphics[scale=.65]{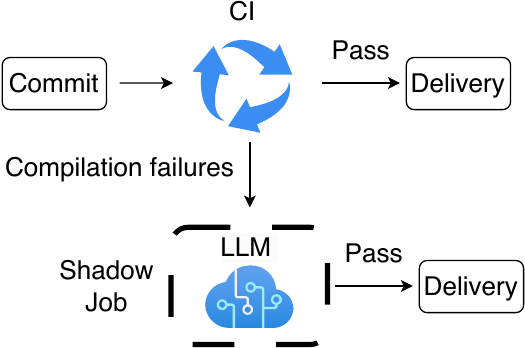}
    \caption{Shadow Job process}
    \label{fig:ShadowJob}
\end{figure} 

We propose a fully automated, generative AI framework for automated repair, incorporating a hardware-in-the-loop system. Our approach analyzes and fixes compilation errors in each CI cycle without disturbing the existing pipeline, using a process we refer to as a~\emph{Shadow Job}. 

Upon a CI failure, the Shadow Job is activated to iteratively determine and apply correction patches until CI is successful (see Fig.~\ref{fig:ShadowJob}). Our Shadow Job requires no additional configuration or alterations to the existing CI setup, thereby avoiding any disruption or complications in the CI configuration. 

\subsubsection{\textbf{LLM CI pipeline}}
Within~\emph{Shadow Job}, we integrate four state-of-the-art LLMs. Fig.~\ref{fig:ShadowJob} shows the workflow of integrating the LLMs into our CI system. When a CI build encounters a compilation failure, an off-the-shelf LLM generates a single-file patch. Subsequently, the shadow job initiates an additional `shadow' CI process to evaluate the effectiveness of the generated patch. If a patch fails to pass the CI testing phase, we repeat the process until a successful patch is found.

\begin{table}
    \caption{Large Language Models}{\small
    \centering
    \begin{adjustbox}{max width=\textwidth}
    \begin{tabular}{lcrc}
    \toprule
    Model & Source code $\subset$ Training set & Parameter Size \\
    \midrule
    CodeT5+  & Yes & 7B \\
    CodeLlama & Yes &7B \\
    Falcon & No &7B  \\
    Bloom & No &7B \\
    \bottomrule
    \end{tabular}
    \end{adjustbox}
    }
    \label{tab:LLMs}
\end{table}

\subsubsection{\textbf{LLM selection}}\label{subsubsec:LLMSelection}
We selected four open-source LLMs (Table~\ref{tab:LLMs}) based on accessibility, code-specific training, dataset diversity, and practical model size. Models must run locally due to confidentiality, be trained on code (e.\,g., CodeT5+, CodeLlama), and include broader data for robustness (e.\,g., Falcon, Bloom). To ensure responsive use in CI repair, we limit model size to 7B parameters for efficient local inference.

\subsubsection{\textbf{Data set preparation}}\label{subsubsec:Data set preparation}
As shown in Fig.~\ref{fig:ShadowJob}, ShadowJob collects commits, source code, and compilation logs from the centralized CI system of a year-long on a highly active \ericsson embedded software project. The dataset includes over~\numprint{40000} C/C++ compilation errors from official CI builds.

We follow guidelines for binary outcome sample size calculations~\cite{fleiss2013statistical}, as outlined in Eq.~\ref{eq:binarySample}, where $p_1$ and $p_2$ are the proportions of the outcome of interest in two different groups being compared. \(\bar{p}\) is the average proportion of both groups. In our study, $p_1$ and $p_2$ represent the CI pass rate after adopting a generated patch for two models, respectively; $n$ represents the number of compilation errors we want to examine.

\begin{equation}\label{eq:binarySample}
\begin{aligned}
n &= \frac{16\,\bar{p}(1-\bar{p})}{(p_1 - p_2)^2}, \quad \text{where} \quad \bar{p} &= \frac{p_1 + p_2}{2}
\end{aligned}
\end{equation}

We ran preliminary trials on each LLM to estimate sample size and found that detecting small differences in pass rates (e.\,g., p$_1$ = 0.25 and p$_2$ = 0.30) requires about \numprint{1000}--\numprint{1200} samples. Based on this, we randomly selected~\numprint{1000} compilation errors from a pool of~\numprint{40000} for our experiments.

\subsection{Methodology of RQ1}
In RQ1, we assess whether an LLM can generate useful fixes for compilation errors, focusing on how prompt design can be optimized to effectively address errors related to dependencies in industrial CI settings.

In APR, using a small code snippet reduces errors or inaccuracies but demands precise syntax. Replacing a large code excerpt with less code may avoid dependency on patch syntax, but may lack full context and may not form a complete syntactic unit within the abstract syntax tree. Balancing these factors is essential for effective program repair. 

To assist AI fixes, we mine older changes (before our APR effort) in the code repository to derive examples of prior successful fixes of given types of compilation errors. In these earlier successfully integrated changes, each compilation error must eventually be followed by at least one successful build with a human fix example. Subsequently, we retrieve the first successful compilation build following the failure. We identify the human fix example by comparing the faulty file with the corresponding file from the successful build. For each compilation error type, we randomly collect one human fix example as an optional prompt item (see below). 

\subsubsection*{Definition of erroneous code and human fix examples}
A \emph{segment} of a program corresponds to one or several lines of code that are not necessarily a syntactic entity.

\emph{Erroneous code} refers to the specific segment of a program that triggers an error, as indicated in the accompanying error message, which typically provides contextual information. 

A \emph{human fix example} refers to a specific segment of a commit that resolves a compilation error. We identify a human fix example by leveraging information from failed build logs, including the commit ID and the location of the faulty line within the affected files. 

\begin{table}
    \caption{Prompt Inputs}{\small
    \centering
    \begin{adjustbox}{max width=\columnwidth}
    \begin{tabular}{@{}llp{4cm}@{}}
        \toprule
        \textbf{ID} & \textbf{Type of Input to the Prompt} & \textbf{Description} \\
        \midrule
        I0 & Source file & The whole file \\
        I1 & Error log & Compilation and CI logs \\
        I2 & Erroneous code snippet & From source file (max 3) \\ 
        I3 & Human fix examples & Human fix by categories \\ 
        \bottomrule
    \end{tabular}
    \end{adjustbox}
    }
    \label{tab:PromptInput}
\end{table}

Table~\ref{tab:PromptInput} outlines the three primary types of inputs used in this study, labeled as \textbf{I0}, \textbf{I1}, \textbf{I2}, and \textbf{I3}.
\begin{itemize}
    \item \textbf{I0} represents the entire source file responsible for the compilation error.
    \item \textbf{I1} contains error logs extracted from the compiler output within the CI system.
    \item \textbf{I2} includes erroneous code snippets extracted from the complete source file.
    \item \textbf{I3} comprises examples of human fixes from historical commits, covering 14 categorized errors types.
\end{itemize}

\begin{table}
    \caption{Prompt Selection}{\small
        \centering
        \begin{adjustbox}{max width=\textwidth}
        \begin{tabular}{cccc}
        \toprule
        {\textbf{No}} & \multicolumn{3}{c}{{\textbf{INPUT COMBINATIONS}}} \\
        \midrule
        0 && Full source file &  \\
        1 &  Error log & &  \\
        2 & & Erroneous code snippet &  \\
        3 & Error log & Erroneous code snippet & \\
        4 & & Erroneous code snippet & Human fix example \\
        5 & Error log & & Human fix example  \\
        6 & Error log & Erroneous code snippet & Human fix example  \\
        \bottomrule
        \end{tabular}
        \end{adjustbox}
        }
    \label{tab:inputcombinations}
\end{table}

Table~\ref{tab:inputcombinations} shows a range of combinations derived from different inputs to generate prompt examples for our study. We have constructed seven distinct prompt examples numbered from No.\,0 to No.\,6 for input into the Shadow Job. Prompt No.\,0 contains the full source file, while prompt No.\,1 includes only the error log. Through experimenting with various combinations, prompt No.\,6 emerges as the most complex one, incorporating the error log, the erroneous code snippet, and the corresponding human fix example. 

\begin{figure}
    \centering  
    \includegraphics[scale=.575]{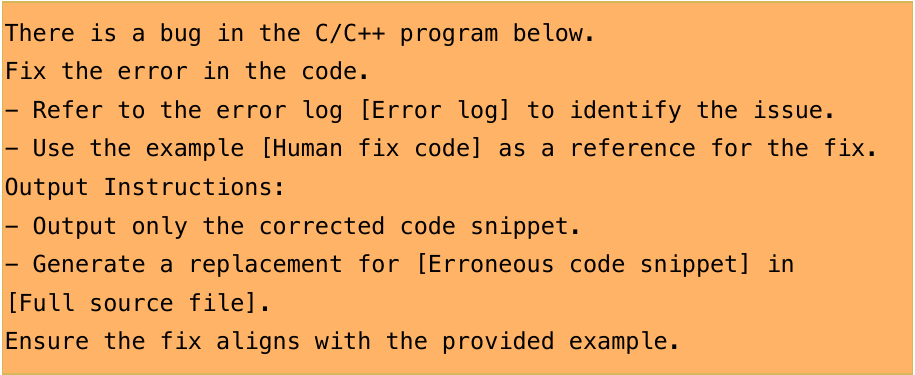}
    \caption{Prompt No.\,6 used to generate patches}
    \label{fig:Prompt6}
\end{figure}

Fig.~\ref{fig:Prompt6} shows the prompt to generate the fix to replace the [Erroneous code snippet] in [Full source file]. 
Configurations consisting solely of human fix examples are deliberately excluded. This exclusion is because inputs comprising only human fix example examples do not provide sufficient information to comprehensively address the compilation errors.

\subsection{Methodology of RQ2}
RQ2 aims to identify which LLMs are most effective for addressing compilation errors. Building on the outcomes of RQ1 that determines the best prompts for LLMs, RQ2 involves conducting systematic experiments across various LLMs.

As shown in Table~\ref{tab:LLMs} and discussed in Section~\ref{subsubsec:LLMSelection}, we select four state-of-the-art LLMs for our experiments. We test each LLM with CI iterations ranging from one to five CI iterations. During these experiments, any instance where no pass verdict was received from the CI was recorded as a failed run. Given the nature of the Shadow Job system we develop, it allows for rapid iteration over multiple runs. Subsequent retries involved the same LLM in the Shadow Job until a successful CI pass was achieved. As part of the evaluation, we measured the Shadow Job pass rate for each model individually to assess their ability to resolve compilation errors.

\subsection{Methodology of RQ3}
In RQ3, we evaluate the operational accuracy and efficiency of our LLM-equipped CI repair system and
focus on the quality of successful repairs. Specifically, we analyze the proportion of human-like (exact match) and plausible fixes among those that pass CI.

Shadow Job performs the initial fix inspection by conducting a case-sensitive match between the LLM-generated patch and the historical fix, ignoring whitespace differences. This is followed by a manual review by two developers with 7 and 10 years of relevant experience. Their similar backgrounds minimized variation in judgment, and minor disagreements allowed us to report average outcomes in the following analysis. Notably, even when a fix matches a historical example, it may not fully align with evolving coding practices—an inherent limitation we acknowledge.

\subsubsection*{Definition of Fix Categories}

We classify fixes into three categories: \emph{exact fix}, \emph{plausible fix}, and \emph{implausible fix}.

An \emph{exact fix} either matches a historical fix or corresponds exactly to how both developers would have written it. 

A \emph{plausible fix} does not meet these criteria but is still judged as reasonable and likely to pass CI. 

An \emph{implausible fix} neither matches historical examples nor is considered valid by the reviewers.

A fix is deemed \emph{reasonable} if it results in a CI pass and is classified as either an exact fix or a plausible fix, i.\,e., it matches a historical fix or is judged plausible by experts.

We applied a consistent experimental setup using optimized prompts across four LLMs, evaluating over~\numprint{1000} compilation error cases. For each successful fix, we measured the time from initial failure to CI pass, allowing up to four Shadow Job iterations. This enabled analysis of time distribution across LLMs and comparison with human-performed fixes.

Previous research~\cite{FuEWEHA24} highlights the significant effort required to resolve compilation errors, often involving small code changes but complex root cause analysis. Thus, we also compare the time efficiency of LLM-based fixes with human intervention, assessing the potential to streamline the CI repair process through faster turnaround and reduced manual effort.

\section{Study Results}\label{sec:studyresults}
\subsection{Results for RQ1}
In RQ1, we examine the impact of different types of contextual information included in the prompt on the LLMs' ability to automatically fix compilation errors. As shown in Table~\ref{tab:PromptSelection}, the first two columns detail the number and different combinations of inputs used to formalize the prompt, as outlined in Section~\ref{sec:methodology}. Columns 3 to 6 illustrate the pass rates for four different LLM-equipped CI systems.

%% Data from full experiments
\begin{table*}[htbp]
    \caption{Prompt Selection}
    \centering
    {\small%
    \begin{tabular}{c c c c c c c c}
    \toprule
    No & \multicolumn{3}{c}{\textbf{Prompt (1 CI iteration)}} & \multicolumn{4}{c}{CI pass rate (\%)} \\
     &  &  &  & \textbf{CodeT5+} & \textbf{CodeLlama} & \textbf{Falcon} & \textbf{Bloom} \\
    \midrule
    0 &  & Full source file &  & 10 & 19 & 16 & \hphantom{0}8 \\
    1 & Log &  &  & 16 & 28 & 13 & 12 \\
    2 &  & Erroneous code &  & 26 & 37 & 24 & 13 \\
    3 & \textbf{Log} & \textbf{Erroneous code} &  & 38 & \textbf{47} & \textbf{38} & 27 \\
    4 &  & \textbf{Erroneous code} & \textbf{Human fix example} & \textbf{44} & 42 & 38 & 25 \\
    5 & Log &  & Human fix example & 43 & 41 & 37 & 34 \\
    6 & \textbf{Log} & \textbf{Erroneous code} & \textbf{Human fix example} & 38 & \textbf{49} & 40 & 38 \\
    \bottomrule
    \end{tabular}%
    }
    \label{tab:PromptSelection}
\end{table*}

For CodeT5+, the highest pass rate (44\,\%) is achieved using a prompt that includes the erroneous code and a manual fix example. In contrast, CodeLlama, Falcon, and Bloom perform best with prompts that additionally include the log (prompt No.\,6) as shown in Table~\ref{tab:PromptSelection}.

These results indicate that more detailed information in the prompt generally leads to better performance for most models. However, the CodeT5+ model performs better with more focused input, suggesting that different models may benefit from different prompt design strategies.

Additionally, some models demonstrate consistently strong performance across multiple prompt configurations, not just the one yielding the highest pass rate. For example, Code\-Llama achieves a 47\,\% pass rate with prompt No.\,3—just 2\,\% below its best. These small variations suggest that certain models are less sensitive to prompt changes, indicating potential robustness in prompt design.

Listing~\ref{ErrorCode} shows a compilation error, where a conditional (\texttt{if}) check was not exhaustive on all types of enumeration \texttt{ObjectType}. This was rejected by a static checker. The corresponding LLM-generated fixing code is shown in Listing~\ref{FixingCode}. The error in the original code snippet is that it only checks if \texttt{type} is \texttt{ObjectType::TYPE\_II} before setting \texttt{dbPrefix} to \texttt{"rx3:"}. The fix changes the condition to include both \texttt{ObjectType::TYPE\_II} and \texttt{ObjectType::TYPE\_I}, ensuring the block executes for these additional types as well. This broader condition likely aligns better with the intended logic of the program. The LLM is able to identify and submit a clean fix, utilizing logical OR (\texttt{||}) for logical operations and short-circuiting. Additionally, the LLM is able to fix static check failures. We provide context by including at least three lines of code (if available) adjacent to the erroneous code. For brevity, this contextual code has been omitted in this document.

\begin{lstlisting}[style=mystyle, caption={Erroneous code 1}, label=ErrorCode]
    // buggy code starts:
    if (type == ObjectType::TYPE_II)
    {        dbPrefix = "rx3:";    }
    // buggy code ends:
\end{lstlisting}
  
\begin{lstlisting}[style=mystyle, caption={Fixing Code 1 (context elided for brevity)}, label=FixingCode]
  -  if (type == ObjectType::TYPE_II)
  +  if ((type == ObjectType::TYPE_II) || (type == ObjectType::TYPE_I))
\end{lstlisting}

This LLM setup can be easily adapted to other industrial production lines with minimal efforts, thanks to its practical and straightforward design. With a similar CI and hardware/software setup, the integration process could be streamlined, achieving optimal results within a single iteration. This approach has demonstrated a maximum success rate of 49\,\%, indicating that the implementation can be effectively replicated across different production lines within the company.

\begin{mdframed}[style=insight, frametitle={Key insight of RQ1:}]
    LLMs perform well in automatically repairing compilation errors, achieving the highest pass rate of 49\,\% in one iteration. It is essential to consider different types of input information in the prompts when generating fixes.
\end{mdframed}

\subsection{Results for RQ2}
In RQ2, we examine which type of model and prompt setup is most effective in addressing compilation errors. Table~\ref{tab:CIiterations-all} lists the performance of all four LLMs with prompt No.\,6. The first column shows the number of CI iterations. 

We achieved a CI pass rate of 63\,\% with the Code\-Llama model. Additionally, with three LLMs—CodeLlama, Falcon, and Bloom—the pass rate remained consistent after three CI iterations, showing no significant change in the fourth and fifth iterations. For LLMs trained without program code, like Falcon and Bloom, we achieved a pass rate of 43\,\%.

\begin{table}
    \caption{CI Iterations – Prompt 3}
    \centering{\small
    \begin{tabular}{c c c c c}
    \toprule
    \textbf{CI iterations} & \multicolumn{4}{c}{CI pass rate (\%)} \\
     & \textbf{CodeT5+} & \textbf{CodeLlama} & \textbf{Falcon} & \textbf{Bloom} \\
    \midrule
    2 & 39 & 51 & 40 & 32 \\
    3 & 39 & 58 & \textbf{42} & \textbf{33} \\
    4 & \textbf{43} & \textbf{59} & 42 & 33 \\
    5 & 43 & 59 & 42 & 33 \\
    \bottomrule
    \end{tabular}}
    \label{tab:CIiterations-3}
\end{table}

Furthermore, we analyzed the performance of different prompts for LLMs with 2 to 5 CI iterations. Since CodeLlama also performs well with prompt No.\,3, Table~\ref{tab:CIiterations-3} lists the CI pass rates of all four LLMs under prompt No.\,3 with 2 to 5 CI iterations. The pass rates for models CodeT5+ and CodeLlama remained consistent after three iterations, with their highest pass rates at 43\,\%, and 59\,\%, respectively. These results are lower than the performance with prompt No.\,6. Falcon and Bloom showed similar results.

We conducted experiments with all four LLMs using prompt No.\,4, as shown in Table~\ref{tab:CIiterations-4}. Surprisingly, the CodeT5+ model with prompt No.\,4 achieve its best performance, with the highest pass rate of 58\,\%, significantly higher than the 43\,\% shown previously.

\begin{mdframed}[style=insight, frametitle={Key insight of RQ2:}]
    We achieved a CI pass rate as high as 63\,\% with Code\-Llama, while CodeT5+ also performed well, reaching a 58\,\% pass rate. LLMs pre-trained with source code demonstrate superior performance in APR for compilation errors compared to models pre-trained without source code.
\end{mdframed}

%% Data from the full experiments
\begin{table}
    \caption{CI Iterations – Prompt 4}
    \centering{\small
    \begin{tabular}{c c c c c}
    \toprule
    \textbf{CI iterations} & \multicolumn{4}{c}{CI pass rate (\%)} \\
     & \textbf{CodeT5+} & \textbf{CodeLlama} & \textbf{Falcon} & \textbf{Bloom} \\
    \midrule
    2 & 48 & 44 & 40 & 28 \\
    3 & 52 & \textbf{55} & 41 & \textbf{29} \\
    4 & \textbf{58} & 55 & \textbf{42} & 29 \\
    5 & 58 & 55 & 42 & 29 \\
    \bottomrule
    \end{tabular}}
    \label{tab:CIiterations-4}
\end{table}

%% Data from the full experiments
\begin{table}
    \caption{CI Iterations – Prompt 6}
    \centering{\small
    \begin{tabular}{c c c c c}
    \toprule
    \textbf{CI iterations} & \multicolumn{4}{c}{CI pass rate (\%)} \\
     & \textbf{CodeT5+} & \textbf{CodeLlama} & \textbf{Falcon} & \textbf{Bloom} \\
    \midrule
    2 & 40 & 58 & 42 & \textbf{39} \\
    3 & 41 & \textbf{63} & \textbf{43} & 39 \\
    4 & 41 & 63 & 43 & 39 \\
    5 & \textbf{45} & 63 & 43 & 39 \\
    \bottomrule
    \end{tabular}}
    \label{tab:CIiterations-all}
\end{table}

\subsection{Results for RQ3}
In RQ3, we explored whether LLMs can generate human-like fixes for compilation errors and examined the time the LLM-equipped Shadow Job takes to resolve such errors. Table~\ref{tab:fixesanalysis} shows the distribution of fixes in three categories: exact fixes, plausible fixes, and implausible fixes, based on the evaluations of two developers.

% Data from the full experiments
\begin{table}
    \centering
    \caption{Fix Analysis}
    {\small
    \begin{tabular}{lccc}
        \toprule
         & \textbf{Exact fixes} & \textbf{Plausible fixes} & \textbf{Implausible fixes} \\
        \midrule
        Developer 1 & 18\hphantom{.0} & 63\hphantom{.0} & 19\hphantom{.0} \\
        Developer 2 & 16\hphantom{.0} & 69\hphantom{.0} & 15\hphantom{.0} \\
        \midrule
        Average     & 17\hphantom{.0} & 66\hphantom{.0} & 17\hphantom{.0} \\ 
        \bottomrule
    \end{tabular}
    }
    \label{tab:fixesanalysis}
\end{table}

We randomly selected 100 samples generated by CodeLlama using prompt No.\,6—which outperformed the other three evaluated LLMs—from the fixes that led to successful CI builds, and distributed them to two \ericsson developers for human analysis. Notably, the two developers' assessments agreed often.

Our analysis reveals that the proportion of exact matches was 17\,\%. The disagreements between developers on exact matches stemmed from the evolution of the software, and one developer disagreed on the suitability of historical changes. The plausible repair rate is 66\,\%, highlighting a substantial potential for LLMs to fix compilation errors. These findings suggest that 83\,\% of the fixes generated by LLMs are reasonable, underscoring their utility in industrial contexts. Some fixes generated by the LLMs may pass the compilation step but fail during subsequent testing phases, which result in a relatively low rate of implausible fixes. 

This discrepancy suggests that while LLMs are proficient at addressing syntax and simple semantic errors, they struggle with complex logical issues that only surface during testing, highlighting the difficulty of addressing deeper logical issues.

\begin{figure}
    \centering  
    \includegraphics[scale=.34]{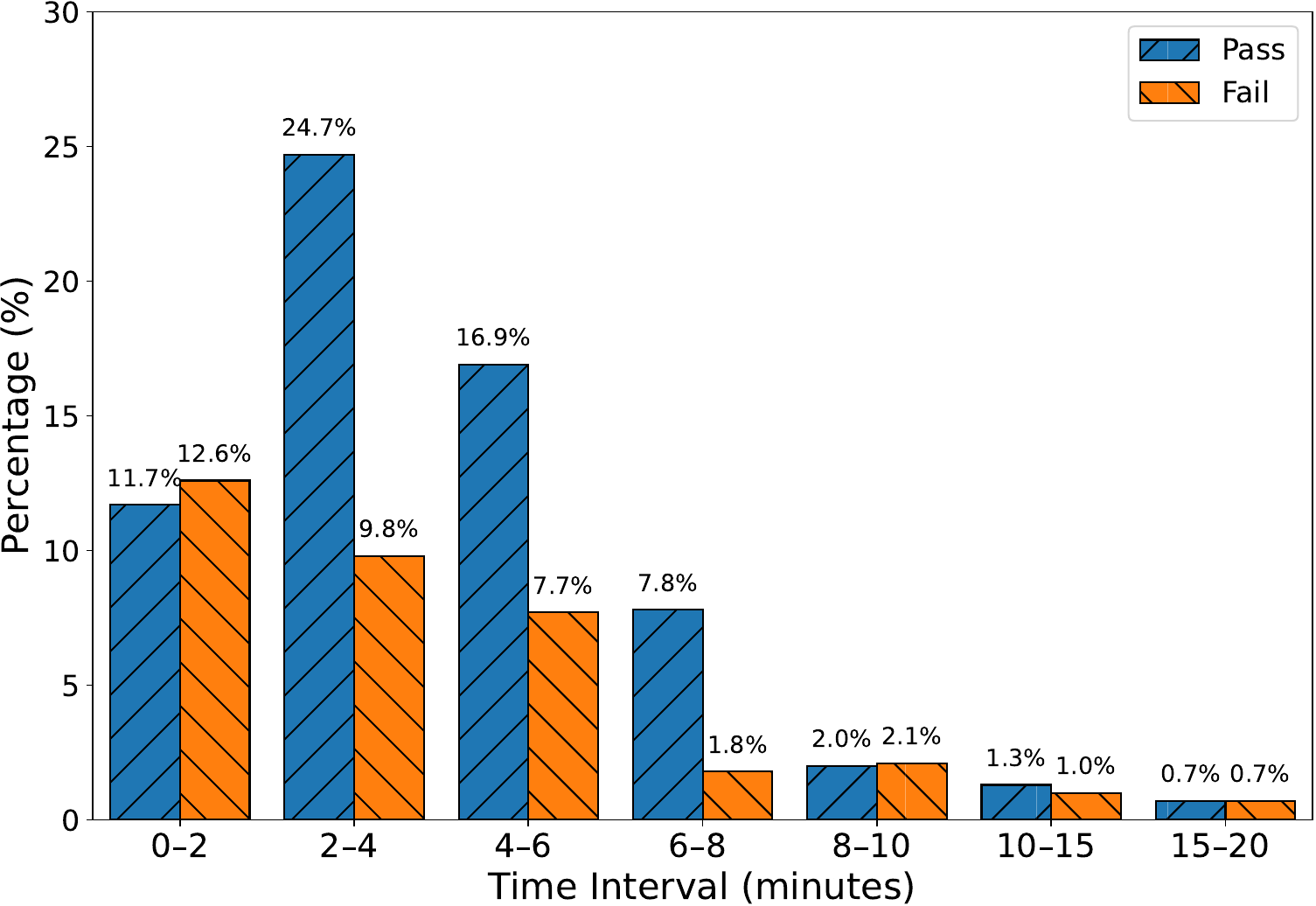}
    \caption{Shadow Job fix time}
    \label{fig:ShadowJobFixTime}
\end{figure}

Fig.~\ref{fig:ShadowJobFixTime} illustrates the time taken for successful and failed builds across all iterations using prompt No.\,6 with CodeLlama in Shadow Job, achieving a 63\,\% pass rate. For time intervals of up to 10 minutes, data is presented in 2-minute increments. Since fewer CI runs exceed 10 minutes, the intervals are expanded to 5 minutes beyond this point.

\begin{mdframed}[style=insight,frametitle={Key insight of RQ3:}]
    LLMs can generate plausible fixes for compilation errors with an 83\,\% success rate, including 17\,\% exact matches, while the LLM-equipped Shadow Job significantly reduces the time required to resolve such errors in CI, achieving a 64\,\% pass rate with the majority of fixes completed within 8 minutes.
\end{mdframed}

Shadow Job efficiently identifies and implements valid fixes, with 11.7\,\% of pass cases completed within 2 minutes, as shown in the first bar. Within 2 to 4 minutes, 24.7\,\% of cases are completed. Nearly 64\,\% of the cases are finished in 8 minutes. This demonstrates that the LLM-equipped Shadow Job is significantly faster than human debugging, greatly reducing the time needed to fix compilation errors. Additionally, 30.1\,\% of failure cases occur within the first 6 minutes, enabling quick feedback and facilitating faster subsequent fixes.

\subsection{Threats to validity}\label{threats2validity}
Our study faces several validity threats. Internally, restricting repairs to single-file changes may overlook cross-file dependencies, while random prompt selection can reduce contextual relevance and affect fix quality. Externally, relying on a single product and untuned LLMs limits generalizability. Construct validity is constrained by using CI pass rate as a proxy for success, which may miss semantic issues, and manual review adds subjectivity. Conclusion validity is limited by sample size and the lack of statistical testing. Future work can address these issues through broader datasets, fine-tuning, semantic evaluation, and statistical analysis.

\section{Discussion}\label{sec:discussion}
Our study demonstrates that LLMs generate high-quality repair suggestions even with limited contextual information. We evaluate four LLMs, selected based on accessibility, code-specific pretraining, dataset diversity, and model size for integration feasibility.

Through experiments on~\numprint{1000} compilation errors in an industrial CI environment, we find that CodeLlama consistently performs best, reaching a peak CI pass rate of 63\,\%. CodeT5+ also performs well (58\,\%), particularly when prompt content is tailored. Falcon, despite not being pre-trained on code, achieves a respectable 43\,\% pass rate after multiple iterations, highlighting the potential of even general-purpose models.

Prompt design plays a critical role in repair effectiveness. Excessive information (e.\,g., full files) lowers performance, with pass rates between 8--19\,\%. Conversely, prompts providing targeted information strike a better balance, helping models focus on the relevant repair context.

Interestingly, good results can still be obtained with simple prompts (e.\,g., log and erroneous code), even without fixing examples. However, the best-performing combinations vary across models, making prompt iteration essential. Code-specific LLMs, such as CodeLlama and CodeT5+, generally benefit more from well-structured prompts.

Multiple CI iterations improve outcomes, especially for more capable models. Performance tends to increase over the initial few iterations, but the improvements gradually level off, indicating diminishing returns with continued retries.

We also observe that 83\,\% of successful fixes are considered reasonable---either matching prior fixes (17\,\%) or judged plausible by human reviewers (66\,\%). Most fixes span just one to four lines, a pattern consistent with previous findings~\cite{FuEWEHA24}. This aligns with many compilation errors, which often stem from localized issues such as missing dependencies, incorrect includes, or syntax mismatches.

For practitioners considering automated compilation error repair, our approach is most effective when fixes are small, self-contained, and limited to a single file—typical of early integration failures in embedded CI. Focused prompts and rapid CI feedback enable quick iteration. It resolves 80\,\% of cases efficiently, and even when it fails, human intervention is needed in under half the cases after an average wait of just eight minutes.

The fast turnaround of compilation jobs—typically accounting for only a small fraction of the total CI time—means that failed repair attempts incur relatively low costs. This efficiency aligns well with our focus on resolving simpler, dependency-related errors. Finally, our Shadow Job architecture enables seamless integration of LLM-based tools into industrial CI systems, allowing for practical, cost-effective experimentation and deployment without disrupting production workflows.

\section{Conclusions and Future Work}\label{sec:conclusion}
Compilation errors in \textit{industrial hardware/software co-development} CI pipelines are common but underexplored in automated program repair~\cite{fu2022prevalence,FuEWEHA24}. Unlike test failures, they offer little runtime context yet are often more straightforward to fix. We address this by integrating state-of-the-art LLMs into a Shadow Job system that enables repeated, non-intrusive repair attempts without altering the main CI pipeline.

Our results show that prompt engineering plays a central role in performance. With carefully structured prompts and multiple iterations, models like CodeLlama can reach up to 63\,\% pass rates, with 83\,\% of the generated fixes judged reasonable. Code-specific pretraining significantly boosts performance, highlighting the importance of using LLMs tuned to the software domain.

Our approach is most effective when errors are localized, fixes are small (typically 1--4 lines), and the relevant context fits within a single file. In fast-build CI environments, the repair loop is efficient---64\,\% of successful fixes are found within the first eight minutes.

Our findings show the effectiveness of LLMs for repairing compilation errors in industrial embedded CI systems. Through lightweight integration, prompt optimization, and iteration, LLM-based repair offers substantial time and resource savings while being compatible with existing workflows.

Future work will focus on improving accuracy and generalizability by fine-tuning models on task-specific datasets tailored to compilation errors in embedded software. We will also refine prompt design to enhance the reliability and usability of generated fixes across diverse environments. Further, ShadowJob’s scope will expand beyond compilation errors to support multi-line, multi-file repairs and integrate testing-phase validation for a more comprehensive CI solution. To strengthen external validity, we will broaden our evaluation to use a diverse set of products, companies, and CI workflows.

\bibliographystyle{ieeetr}
\balance
\bibliography{references.bib}
\end{document}